\documentclass[a4paper,fleqn]{cas-dc}

\usepackage[numbers]{natbib}
\usepackage{algorithm}
\usepackage{algpseudocode}
\usepackage{algorithmicx}

\usepackage{graphicx}            
\usepackage{caption} 
\usepackage{subcaption}
\usepackage{hyperref} 
\usepackage{cleveref} 

\newtheorem{example}{Example}

\begin{document}
\let\WriteBookmarks\relax
\def\floatpagepagefraction{1}
\def\textpagefraction{.001}
\shorttitle{Bayesian thinning algorithm}
\shortauthors{Z. L. Deng et~al.}

\title [mode = title]{A Bayesian thinning algorithm for the point source identification of heat equation}                      



\author[1]{Zhiliang Deng}[
                        ]
\cormark[1]
\fnmark[1]
\ead{dengzhl@uestc.edu.cn}

\credit{Conceptualization of this study, Methodology, Software}

\affiliation[1]{organization={School of Mathematical Sciences, University of Electronic Science and Technology of China},
                addressline={Xiyuan Ave}, 
                city={Chengdu},
                postcode={611731}, 
                state={Sichuan},
                country={China}}

\author[1]{Chen Li}[
]

%
%

\author[2]{Xiaomei Yang}
\cormark[2]
\fnmark[1]
\ead{yangxiaomath@swjtu.edu.cn}

\affiliation[2]{organization={School of Mathematics, Southwest Jiaotong University},
                addressline={Xipu Campus}, 
                city={Chengdu},
                postcode={610097}, 
                state={Sichuan}, 
                country={China}}

\cortext[cor1]{Zhiliang Deng}
\cortext[cor2]{Xiaomei Yang}
%

\begin{abstract}
In this work, we propose a Bayesian thinning algorithm for recovering weighted point source functions in the heat equation from  boundary flux observations.
The major challenge in the classical Bayesian framework lies in  constructing suitable priors for such highly structured unknowns.
To address this, we introduce a level set representation on a discretized mesh for the unknown, which enables  the  infinite-dimensional Bayesian framework to the reconstruction.
From another perspective, the point source configuration can be modeled as a marked Poisson point process (PPP), then a thinning mechanism is employed to selectively retain points.
These two proposals are complementary with the Bayesian level set sampling generating candidate point sources and the thinning process acting as a filter to refine them.
This combined framework is validated through numerical experiments, which demonstrate its accuracy in reconstructing point sources.


\end{abstract}



\begin{keywords}
inverse source problem \sep point source \sep Bayesian approach \sep Poisson point process \sep level set
\end{keywords}

\maketitle

\section{Introduction}

In this paper, we investigate the inverse problem of identifying a point source term  in the heat equation within a bounded domain \( \Omega \subset \mathbb{R}^2 \). The problem is formulated as follows:
\begin{align}
\begin{cases}\label{eq1}
u_t - \Delta u = \sum_{i=1}^{N} w_i \delta_{x^{(i)}}, & x \in \Omega, \, t > 0, \\
u(x, 0) = 0, & x \in \Omega, \\
u(x, t) = 0, & x \in \partial \Omega, \, t > 0,
\end{cases}
\end{align}
where \( x^{(i)} \) denotes the location of the \( i \)-th point source, and \( w_i \) represents its corresponding intensity. By $f$ denote $\sum_{i=1}^{N} w_i \delta_{x^{(i)}}$. The problem is studied under the framework of recovering \( f \) from observational data of the normal derivative \( \partial u/\partial n \) on the boundary \( \partial \Omega \).

The inverse problem of point source identification arises in a wide range of scientific and engineering disciplines. Notably, it plays a critical role in (i) environmental engineering, where it is employed for contaminant source localization \cite{Badia2005,Egan1972,Kovalets2011}, and (ii) biomedical imaging, particularly in EEG-based neural source reconstruction \cite{Badia2000, Baratchart2005}. More broadly, such inverse problems emerge in contexts such as heat transfer, wave propagation, and electrostatics.
A considerable body of literature has addressed both the theoretical identifiability of point sources and the development of stable numerical algorithms across different types of partial differential equations (PDEs). For example, Ling et al. \cite{Ling2006, Ling2009} investigate the minimal number of measurements of the state variable $u$ needed to uniquely determine source locations, quantities, and intensities in heat equations. They propose a stable numerical approach based on the method of fundamental solutions combined with collocation techniques. This framework is further extended by Gu et al. \cite{Gu2025}, who incorporate flux measurements $\partial u/\partial n$ in bounded domains and address the resulting inverse problem using an optimization-based strategy.
For elliptic PDEs, Badia et al. \cite{Badia2000} study the identifiability of point sources from boundary data and introduce an algebraic reconstruction algorithm. Baratchart et al. \cite{Baratchart2005} build upon this work by reformulating the inverse problem as one of best rational or meromorphic approximation on the boundary, offering a rigorous analytical foundation. In the context of one-dimensional advection–dispersion–reaction equations, Badia et al. \cite{Badia2005} establish identifiability conditions for both spatial source locations and time-dependent source intensities. For wave equations in one spatial dimension, Bruckner~~\cite{Bruckner2000} analyzes stability properties and proposes two provably stable numerical schemes.
More recent advances include the works of Faria et al. \cite{Faria2022_c, Faria2022_s}, who investigate inverse problems governed by the modified Helmholtz and Poisson equations. In their approach, the forward problems are solved using the method of fundamental solutions, while the inverse problems are formulated as optimization problems. In the heterogeneous Helmholtz setting, Ren et al. \cite{Ren2019} derive stability estimates and propose a projection-based numerical reconstruction method. Zhang et al. \cite{Zhang2024} consider the identification of both acoustic point sources and obstacles in two-dimensional Helmholtz systems using Cauchy data.

These prior studies motivate the development of a flexible Bayesian framework for point source identification, as explored in this paper. Despite the rich body of literature on optimization-based techniques, existing numerical approaches for point source reconstruction remain predominantly grounded in conventional deterministic frameworks, with limited methodological innovation. This is particularly striking given the proven success of Bayesian methods in tackling a wide range of inverse problems \cite{Dashti2017, Latz2020, Latz2023, Reich2021, Stuart2010}, which have not yet been systematically applied to this class of reconstruction problems.
One of the main challenges in adopting a Bayesian perspective lies in the appropriate specification of prior distributions, especially considering the discrete, spatially localized nature of point sources. To address this, we propose a novel Bayesian framework for identifying point sources in the heat equation, implemented through a two-stage inference procedure.
The first stage constructs a prior over the unknown source locations using Gaussian random fields combined with a level set representation. This formulation enables smooth, flexible modeling of source geometry while allowing for efficient posterior sampling via the preconditioned Crank–Nicolson (pCN) algorithm. The resulting samples provide a set of candidate source locations.
In the second stage, these candidates are reinterpreted as realizations from a marked Poisson point process (PPP), upon which we apply a leave-one-out thinning algorithm to remove redundant or spurious sources. This thinning procedure preserves the spatial structure of the field while enhancing sparsity and interpretability. The two-stage correction process is iteratively applied to refine both the source locations and the corresponding intensities.
Overall, the proposed method offers a principled and fully Bayesian approach to point source reconstruction, delivering robust and interpretable estimates of both source geometry and strength.

The remainder of this paper is organized as follows. In section 2 the sketch of Bayesian approach is explained. We then introduce the level set parameterization to characterize the target variate. We reformulate the level set parameterization as a marked PPP in section 3. In section 4, we propose an iterative process which combines the pCN algorithm and the thinning process. By extensively numerical experiments, we verify the numerical effectiveness. The paper concludes with a summary in section 5. 

\section{Bayesian approach for the point source reconstruction}
\subsection{The Bayesian approach}
This section presents the fundamental principles of Bayesian methodology for point source reconstruction. 
%
Within this Bayesian framework, our objective is to determine the posterior distribution of the source term \( f \) conditioned on the observed Neumann boundary data \( \partial u/\partial n\). Let \( g \) represent the noisy measurement data, modeled as:
\begin{align}
g = {\partial u}/{\partial n} + \eta,
\end{align}
where \( \eta\sim N(0, \Xi) \) denotes the measurement noise. In inverse problems terminology, $f\rightarrow g$ is the forward map, 
\begin{align}
\mathcal{K}(f)=g,
\end{align}
and the inverse problem is to recover $f$ given $g$.
 We denote the conditional random variable by \( f | g \), with its posterior distribution represented as \( \mu(f | g) \). 
The unknowns include the number $N$, the locations $x^{(i)}$, and the intensities $\lambda_i$. Therefore, we may consider the posterior distribution in a finite-dimensional space $\mathbb{R}^d$. 
Following Bayes' formulation, the posterior distribution can be expressed as
\begin{align}
\mu(f|g)={\exp(-{|\Xi^{-1}(\mathcal{K}f-g)|^2}/{2})\mu_0(f)}/{Z},
\end{align}
where $Z:=\int\exp(-{|\Xi^{-1}(\mathcal{K}f-g)|^2}/{2})d\mu_0(f)$ is the normalization constant ensuring $\mu(f|g)$ is a valid probability distribution. 
The corresponding probability density satisfies
\begin{align}
p(f|g)\propto\exp\left(-\Phi(f; g)\right)p_0(f),
\end{align}
where $\Phi(f; g)$ is the negative log-likelihood (potential energy term).
The term $\mu_0(f)$ (resp. $p_0(f)$) is  the prior distribution (resp. prior density),  encoding assumptions about $f$ before observing data. Moreover, we define the total energy function by $V(f):=\Phi(f; g)-\log p_0(f)+\log Z$. 



To sample from this posterior, a powerful perspective arises by viewing $\mu(f|g)$ as the invariant (equilibrium) distribution of a stochastic process. Specifically, we consider a Langevin-type dynamics, which describes the evolution of a fictitious particle in the state space of $f$ under the influence of both the gradient of the negative log-posterior (i.e., drift) and stochastic noise:

$$
df_t = -\nabla V(f_t)\,dt + \sqrt{2}\,dW_t.
$$
Here, $W_t$ is a Brownian motion, and the dynamics are designed so that the posterior $\mu$ is invariant under this flow.
The associated Fokker–Planck equation governs the time evolution of the probability density $p(f, t)$ of $f_t$:
$$
{\partial p}/{\partial t} = \nabla \cdot \left(p\, \nabla V(f) + \nabla p \right).
$$
This partial differential equation describes how the distribution of particles evolves under the Langevin dynamics. Crucially, the posterior $p(f|g)=e^{-V}$ is a stationary solution of this equation.

%
%

The Bayesian posterior thus admits a dual interpretation: dynamically, as the equilibrium distribution of a stochastic differential equation (SDE), and statistically, as the conditional probability distribution of the unknown $f$ given observations $g$. This dual perspective establishes a foundational link between stochastic sampling methods—such as Langevin MCMC and the Metropolis-adjusted Langevin algorithm (MALA)—and partial differential equation (PDE) formulations \cite{Cotter2013, Dashti2017, Stuart2010}.

The strength of this framework becomes particularly evident in the context of infinite-dimensional inverse problems, especially those constrained by PDEs. In such settings, naive extensions of finite-dimensional MCMC algorithms often fail to produce well-defined or efficient samplers. This has motivated the development of geometrically informed algorithms that remain valid and robust in infinite-dimensional spaces. 
%
%
We now present the canonical form of the pCN algorithm (Algorithm \ref{alg1}), which is used in this paper.
\begin{algorithm}
\begin{algorithmic}[1]
\State 
Given an initial state $f^{(0)}$.  For $n=0, 1, \cdots$, 
\State Propose a move
$$
\hat{f}^{(n)}=\sqrt{1 - \beta^2}\, f^{(n)} + \beta\, \xi, \quad \xi \sim \mu_0.
$$
\State
Then accept the proposal with probability
$$
a(f^{(n)}, \hat{f}^{(n)}) = \min\left\{1, \exp\left( \Phi(f^{(n)}; g) - \Phi(\hat{f}^{(n)}; g) \right) \right\},
$$
otherwise set $f^{(n+1)} = f^{(n)}$.

\end{algorithmic}
\caption{The pCN algorithm}
\label{alg1}
\end{algorithm}


\subsection{Level set parameterization}



In our problem, the unknown field consists of a collection of localized point sources, resulting in a highly sparse and discontinuous structure. Directly estimating such fields within a Bayesian framework is particularly challenging due to their non-Gaussian nature, lack of smoothness, and the poor regularization behavior of conventional priors. Similar challenges arise in applications such as source identification in electrostatics, subsurface detection, and inverse scattering.

To overcome these limitations, the level set approach introduces an auxiliary function $\phi : \Omega \to \mathbb{R}$, called the level set function, which implicitly encodes the geometry of the unknown structure. 
First, we generate a sufficiently fine mesh over the domain $\Omega$ and collect the mesh nodes as $\{x^{(i)}\}$.
The physical parameter $f$ is then defined through a threshold-based transformation of a latent random field $\phi$ as follows:
\begin{align}\label{eqn6}
f(x) = F(\phi(x)):= \sum\nolimits_{\phi(x^{(i)}) > c} \phi(x^{(i)}) \delta_{x^{(i)}},
\end{align}
where $c>0$ is a prescribed threshold.
This reparameterization transforms the original geometric inverse problem into an inference problem on the underlying field $\phi$, which can now be assigned a smooth Gaussian prior, e.g., $\phi \sim \mathcal{N}(0, \mathcal{C})$, defined over a suitable Hilbert space.
As a result, $f$ can be interpreted as a truncated Gaussian point field, represented by $\mathcal{N}(0, \mathcal{C}) \cdot \mathbb{I}_{\phi(x) > c}$, evaluated only at the discrete mesh locations $\{x^{(i)}\}$. It can be observed that the conditional field values $\{\phi(x^{(i)}) \mid \phi(x^{(i)}) > c\}$ still follow a Gaussian distribution, with covariance matrix $C$, where $C$ is the discretization of the covariance operator $\mathcal{C}$ restricted to the subset of mesh points satisfying $\phi(x^{(i)}) > c$.

The characterization in \eqref{eqn6}  targets scenarios where the number, locations and intensities  are all unknown. However, the framework can also be applied to the case where only the number and the spatial locations  are uncertain and each point source is of the same intensity. For simplicity, we assume that the true source is of this form
\begin{align}\label{const}
f(x)=\sum\nolimits_{i=1}^N \delta_{x^{(i)}},
\end{align}
i.e., $w_i=1$ for $i=1, 2, \cdots, N$. For this case, this corresponding level set parameterization as in \eqref{eqn6} is modified as 
\begin{align}\label{const_p}
f(x)=F(\phi(x)):= \sum\nolimits_{\phi(x^{(i)}) > c} \delta_{x^{(i)}}.
\end{align}


In the Bayesian setting, the posterior distribution of $\phi$ given data $g$ is:
$$
\frac{d\mu(\phi|g)}{d\mu_0(\phi)} \propto \exp\left(-\Phi(F(\phi); g)\right),
$$
where  $\mu_0$ denotes the Gaussian prior on $\phi$. This formulation benefits from the well-posedness properties established for Gaussian priors in infinite-dimensional Bayesian inverse problems, thereby enabling stable and consistent inference  \cite{Iglesias2016}. The regularity imposed by the Gaussian prior also helps control the complexity of the level set function $\phi$.

%

Sampling from the posterior distribution $\mu(\phi|g)$ can be performed efficiently using function-space Markov chain Monte Carlo (MCMC) methods, such as the preconditioned Crank–Nicolson (pCN) algorithm. The level set parameterization introduces a smooth latent field $\phi$, through which the geometry of the interface is implicitly represented. This smoothness, together with the function-space formulation, allows MCMC samplers to explore the posterior distribution in an infinite-dimensional setting without suffering from mesh dependence or the curse of dimensionality.
When combined with geometrically informed MCMC algorithms, the level set approach provides a rigorous and flexible framework for Bayesian inference in geometric inverse problems. This methodology has been successfully applied in various PDE-constrained inverse problems, demonstrating its robustness and adaptability \cite{Huang2021, Iglesias2016}.
The theoretical well-posedness of the posterior and the dimension-independent performance of function-space MCMC methods ensure reliable uncertainty quantification in complex geometric settings.

\section{Poisson point process}

The spatial distribution of point sources can be modeled as a point process. Formally, a point process is a random collection of points in a measurable metric space $(\mathbb{X}, \mathcal{X})$, which can be characterized as a random counting measure $\mathfrak{N}$ such that: 
For a measurable set $A$, $\mathfrak{N}(A)$ counts the (random) number of points falling in $A$. A Poisson point process (PPP) is fundamental model where:
\begin{description}
\item[a)] Counts in disjoint regions are independent.
\item[b)] The number of points in $A$ follows a Poisson distribution:
\begin{align}
\mathfrak{N}(A)\sim Po(\Lambda(A)),
\end{align} 
where $\Lambda(A)$ is the intensity measure (mean number of points in $A$).
\end{description}
We assume that the intensity measure $\Lambda$ has a density with respect to the Lebesgue measure, denoted by $\lambda$, which is referred to as the intensity function. That is, for any measurable set $A \subseteq \mathbb{R}^d$,
$$
\Lambda(A) = \int_A \lambda(x)\, dx.
$$
For a homogeneous PPP in $\mathbb{R}^d$, $\Lambda(A)=\lambda|A|$ ($\lambda$: constant intensity, $|A|$: volume of $A$). 



There are several standard methods for constructing new Poisson point processes (PPPs) from existing ones. Among them, two widely used techniques that modify a single PPP are thinning, which randomly retains points according to a specified retention probability, and marking, which augments each point with a random mark drawn from a given distribution, yielding a marked PPP. In addition to these operations on a single process, another fundamental construction is the superposition of multiple independent PPPs, which results in a new PPP with an intensity equal to the sum of the individual intensities.


%
%
%
%

 In the classical independent thinning framework, each point $x \in S$ is independently retained with probability $t(x) \in [0, 1]$, where $S$ is a realization of $\mathfrak{N}$ and $t(x)$ is a measurable thinning function. The resulting process $\mathfrak{N}'$ is again a PPP, with a reduced intensity $t(x)\lambda(x)$.
 
Let $(\mathbb{W}, \mathcal{W})$ be a measurable space.  In a marked PPP, each point $x \in S$ is independently assigned a random mark $m \in \mathbb{W}$ drawn from a mark distribution $\kappa(x, m)$, where 
$\kappa(\cdot, \cdot)$ is a probability kernel, i.e., $\kappa: \mathbb{X}\times\mathbb{W}\rightarrow [0, 1]$, $\kappa(x, \cdot)$ is a probability measure for each $x\in\mathbb{X}$ and $\kappa(\cdot, C)$ is measurable for each $C\in\mathcal{W}$. 
 We call $(\mathbb{W}, \mathcal{W})$  the mark space. 
 The outcome is a marked point process $\tilde{\mathfrak{N}}$ on the product space $\mathbb{X} \times \mathbb{W}$. 
 If the original process $\mathfrak{N}$ is a PPP and the marks are conditionally independent given locations, then the marked process $\tilde{\mathfrak{N}}$ is also a PPP, with an intensity measure given by $\Lambda(dx) \kappa(dx, dm)$.
 
We can interpret \eqref{eqn6} in the framework of a marked PPP.  Here, the object variates $w_i$ and $x^{(i)}$ in \eqref{eq1} are treated as marked points $(x^{(i)}, w_i)$, where $S=\{x^{(i)}\}$ is a realization of a PPP and $w_i$ serves as the associated mark. 
Accordingly, \eqref{const_p} corresponds to the case of PPP, where just the locations and number of point sources is to be determined.





\section{Bayesian thinning algorithm and numerical tests}

\subsection{Bayesian thinning algorithm}
As previously discussed, the level set method enables us to tackle the point source reconstruction problem within the framework of infinite-dimensional Bayesian inversion. Although effective for smooth interfaces, this approach may struggle with highly discontinuous or fractal-like sources without specialized regularization. Moreover, given the extreme sparsity of discrete point sources, relying solely on this framework may lead to computational inefficiency.
From the perspective of PPP, the reconstruction target is modeled as a marked point process. While thinning provides a methodology for obtaining approximate reconstructions, it requires evaluating the retention or removal probability for each point. This leads to two critical limitations: (1) computational complexity scales with the number of points, and (2) true points may be erroneously deleted (false negatives), particularly problematic in high-density scenarios.

In this subsection, we propose a Bayesian thinning algorithm that combines the pCN method with a thinning procedure. 
First, finite pCN updates are applied to the level set function, producing candidate point sources. Then, a thinning procedure removes redundant or insignificant sources. This cycle is repeated iteratively.
The procedure is summarized in Algorithm~\ref{alg2}.
\begin{algorithm}
\begin{algorithmic}[1]

\State  Set the maximum iteration step $N_{\max}$, the threshold parameter $c$.  Give an initial state $\phi$ and get the corresponding initial source $f$. 
\State While the iteration step $n\leq N_{\max}$, implement the following two-step procedure:

{\bf Step 1 (Level set update):} 

By finite steps, using Algorithm \ref{alg1} to update the level set function $\phi$ and generate a candidate point source function $f$ according to \eqref{eqn6}, which includes the locations $\{x^{(i)}\}_{i=1}^J$ and the corresponding intensities $\{w_i\}_{i=1}^J$.
Denote $\{(x^{(i)}, w_i)\}_{i=1}^J$ by $\theta$.

{\bf Step 2 (The thinning procedure):}

   For each $j = 1$ to $J$, denote the leave-one-out set by $\breve{\theta}_j:=\{(\breve{x}^{(i)}, \breve{w}_i)\}_{i\neq j}$, and the corresponding source function by 
   \begin{align}
   \breve{f}_j=\sum\nolimits_{i\neq j} w_i\delta_{x^{(i)}}
   \end{align}
    and evaluate its probability by
        \begin{align}
        \alpha=\min\left\{1, {p(\breve{f}_j|g)}/{p(f|g)}\right\}.
        \end{align}
         \begin{description}
        \item[\hspace{4em}-] If $\alpha>\text{Uniform}(0, 1)$, we remove the point $(x^{(j)}, w_j)$, otherwise retain it.
         
         \item[\hspace{4em}-] If the point $(x^{(j)}, w_j)$ is deleted from the candidate point sources, we remove $\{(x^{(j)}, w_j)\}$ from $\theta$ and update  $f$ by removing the point source, i.e., 
         $$f=\sum\nolimits_{i\neq j}w_i \delta_{x^{(i)}}.$$
         \end{description}


\end{algorithmic}
\caption{The Bayesian thinning algorithm}
\label{alg2}
\end{algorithm}

\subsection{Numerical experiments and discussions}

In this subsection, we present some numerical examples to demonstrate the effectiveness of the proposed method. We generate data by adding the relative Gaussian noise to numerical solution as
\begin{align}
g=\mathcal{K}(f)+\|\mathcal{K}(f)\| \delta\xi,
\end{align}
where $\delta>0$ is the noise level, $\xi$ is a  normal Gaussian white noise. For all examples, we fix $\delta=1\%$.

\begin{example}\label{exam1}
We first examine cases with data collected at discretized time points $t_i = i\Delta t$ for $i = 1, \cdots, 1/\Delta t$, where $\Delta t = 0.01$. The configurations of a single observation location and two observation locations are considered, with the point source intensity held constant at $1$.
\end{example}

\begin{figure}[htbp]
    \centering
    \begin{subfigure}[b]{0.2\textwidth}
        \includegraphics[width=\textwidth]{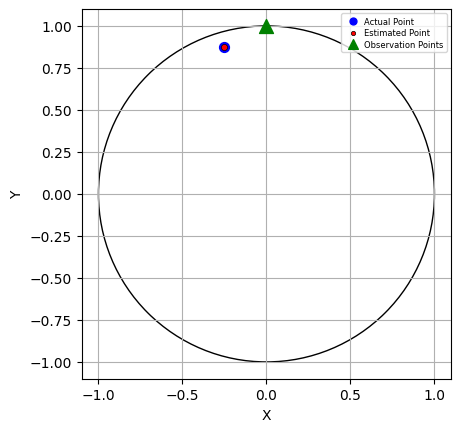}
        \caption{}
        \label{fig:subfig1}
    \end{subfigure}
    \begin{subfigure}[b]{0.2\textwidth}
        \includegraphics[width=\textwidth]{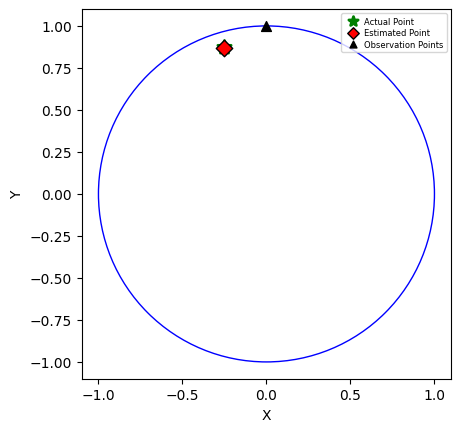}
        \caption{}
        \label{fig:subfig2}
    \end{subfigure}
    \begin{subfigure}[b]{0.2\textwidth}
        \includegraphics[width=\textwidth]{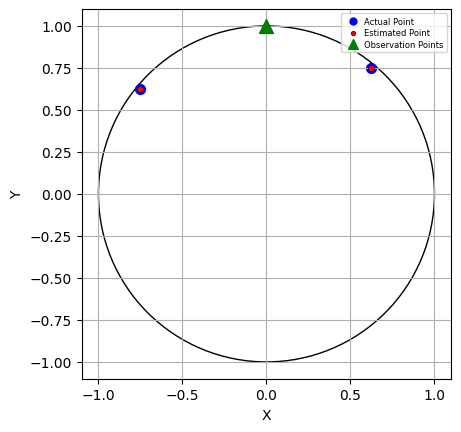}
        \caption{}
        \label{fig:subfig3}
    \end{subfigure}
    \begin{subfigure}[b]{0.2\textwidth}
        \includegraphics[width=\textwidth]{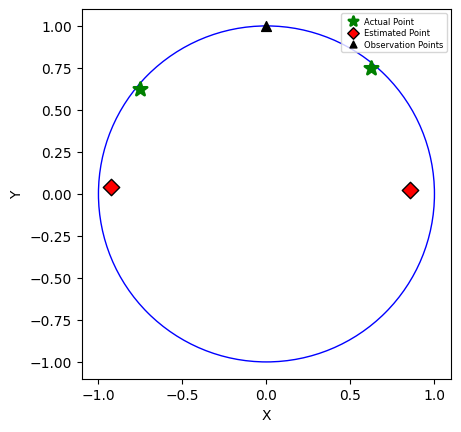}
        \caption{}
        \label{fig:subfig4}
    \end{subfigure}
    \caption{Comparison of reconstruction results using data from a single observation location. 
    The proposed method ((a) $\&$ (c)) is compared against the method of \cite{Gu2025} ((b) $\&$ (d)).}
    \label{fig1}
\end{figure}

\begin{figure}[htbp]
    \centering
    \begin{subfigure}[b]{0.2\textwidth}
        \includegraphics[width=\textwidth]{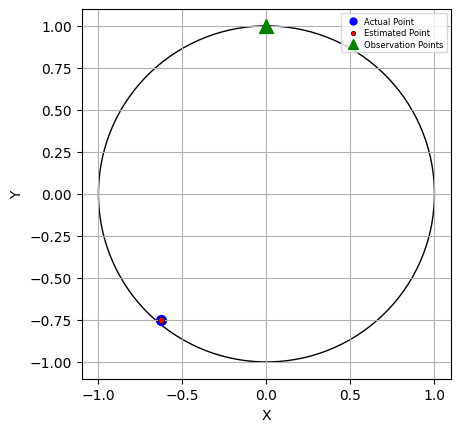}
        \caption{$N=1$}
        \label{fig2_a}
    \end{subfigure}
    \begin{subfigure}[b]{0.2\textwidth}
        \includegraphics[width=\textwidth]{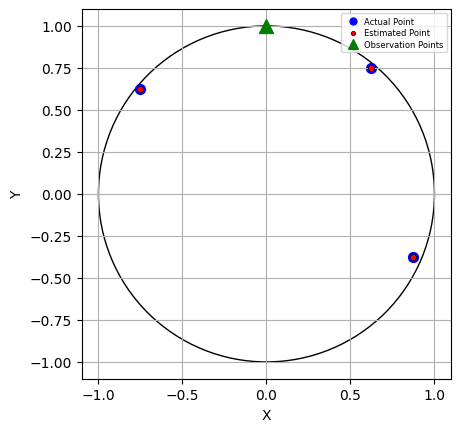}
        \caption{$N=3$}
        \label{fig2_b}
    \end{subfigure}
    \caption{Reconstruction results using data collected at a single observation location during the time interval $0<t<1$.}
    \label{fig2}
\end{figure}

\begin{figure}[htbp]
    \centering
    \begin{subfigure}[b]{0.2\textwidth}
        \includegraphics[width=\textwidth]{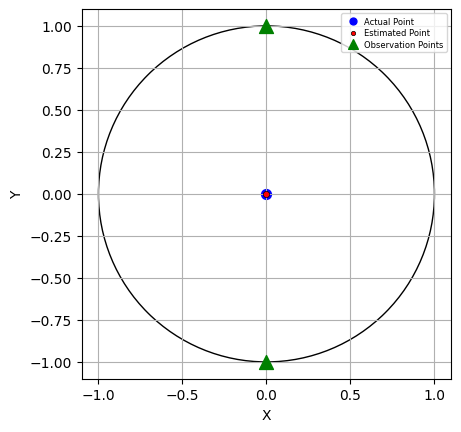}
        \caption{$N=1$}
        \label{fig3_a}
    \end{subfigure}
    \begin{subfigure}[b]{0.2\textwidth}
        \includegraphics[width=\textwidth]{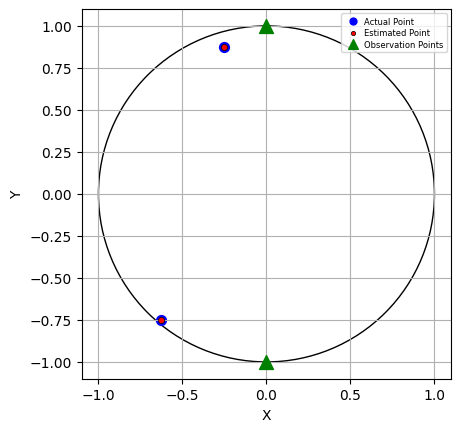}
        \caption{$N=2$}
        \label{fig3_b}
    \end{subfigure}
    \begin{subfigure}[b]{0.2\textwidth}
        \includegraphics[width=\textwidth]{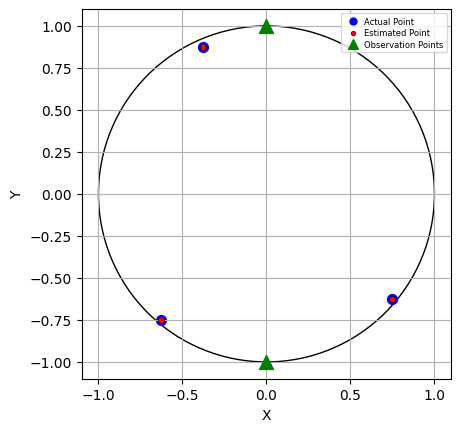}
        \caption{$N=3$}
        \label{fig3_c}
    \end{subfigure}
    \begin{subfigure}[b]{0.2\textwidth}
        \includegraphics[width=\textwidth]{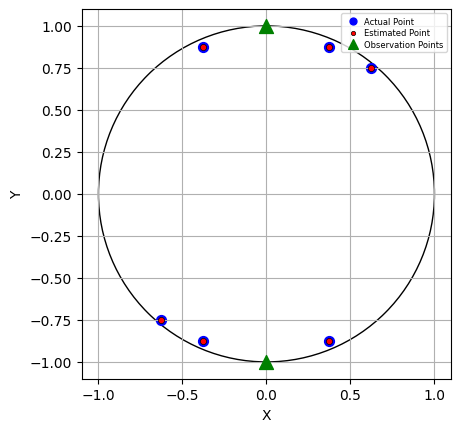}
        \caption{$N=6$}
        \label{fig3_d}
    \end{subfigure}
    \caption{Reconstruction results using data collected at two observation locations for fixed $t=1$.}
    \label{fig3}
\end{figure}

First, we test the numerical effectiveness using a single observation location.  \cite{Gu2025} proposes an iterative scheme for the reconstruction problem. Fig.~\ref{fig1} compares the numerical performance of our method against that of \cite{Gu2025} for the case of a single point source and multiple point sources. The results demonstrate that our method is effective and achieves performance competitive with that of \cite{Gu2025}. 
Fig.~\ref{fig2} displays reconstruction results for two cases: (a) a single point source away from the observation point, and (b) three point sources. We then check the cases of $N=1, 2, 3, 6$ using two observation locations and display the reconstruction results  in Fig.~\ref{fig3}.  
In all cases, the proposed method accurately reconstructs the source locations.  

%
%

\begin{figure}[htbp]
    \centering
    \begin{subfigure}[b]{0.15\textwidth}
        \includegraphics[width=\textwidth]{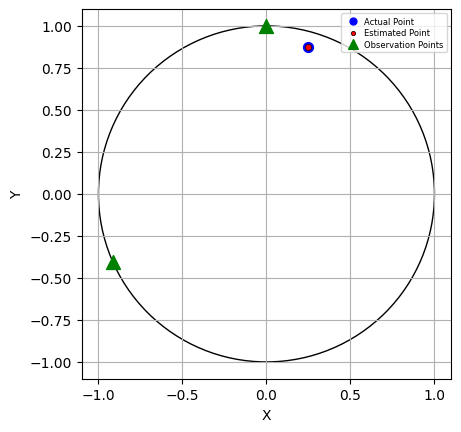}
        \caption{$N=1$}
        \label{fig4_a}
    \end{subfigure}
    \begin{subfigure}[b]{0.15\textwidth}
        \includegraphics[width=\textwidth]{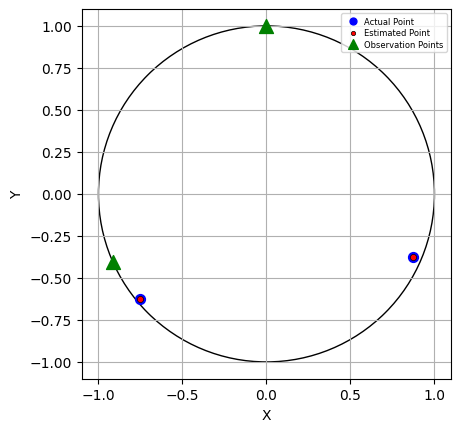}
        \caption{$N=2$}
        \label{fig4_b}
    \end{subfigure}
    \begin{subfigure}[b]{0.15\textwidth}
        \includegraphics[width=\textwidth]{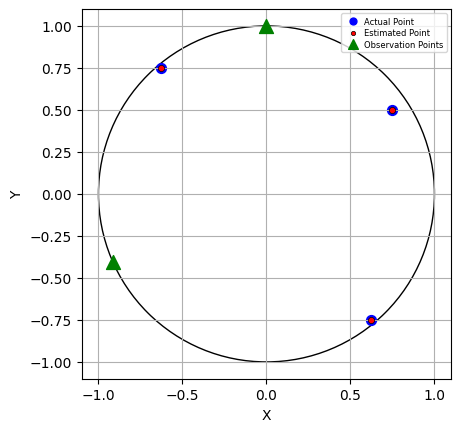}
        \caption{$N=3$}
        \label{fig4_c}
    \end{subfigure}
    \caption{Reconstruction results using data collected at two observation locations for fixed $t=1$ with the thinning process.}
    \label{fig4}
\end{figure}

\begin{figure}[htbp]
    \centering
    \begin{subfigure}[b]{0.2\textwidth}
        \includegraphics[width=\textwidth]{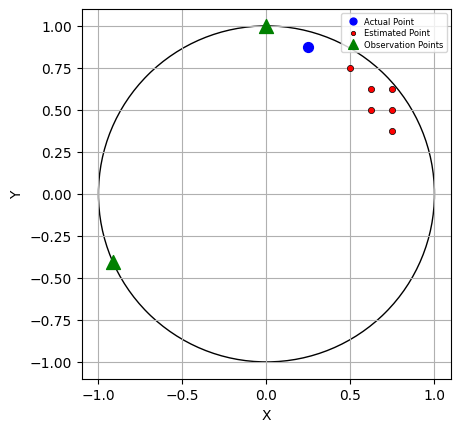}
        \caption{$N=1$}
        \label{fig5_a}
    \end{subfigure}
    \begin{subfigure}[b]{0.2\textwidth}
        \includegraphics[width=\textwidth]{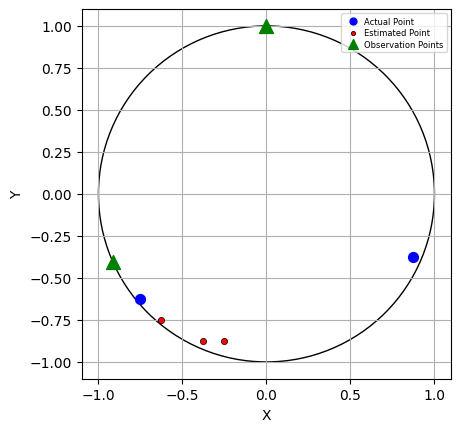}
        \caption{$N=2$}
        \label{fig5_b}
    \end{subfigure}
    \caption{Reconstruction results using data collected at two observation locations during the time interval $0<t<1$ when the thinning process is removed.}
    \label{fig5}
\end{figure}

\begin{figure}[htbp]
    \centering
    \begin{subfigure}[b]{0.22\textwidth}
        \includegraphics[width=\textwidth]{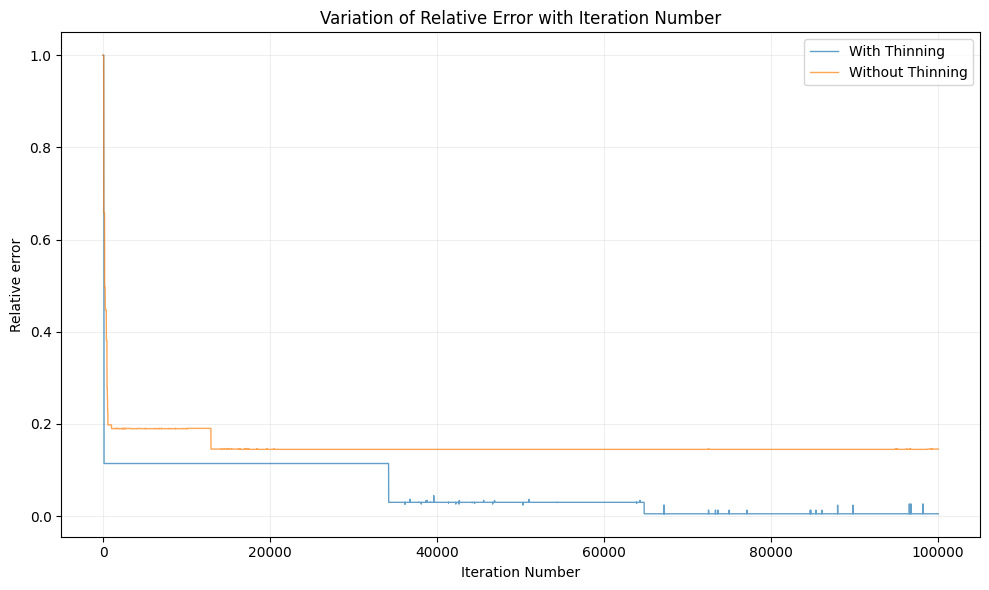}
        \caption{$N=1$}
        \label{fig6_a}
    \end{subfigure}
    \begin{subfigure}[b]{0.22\textwidth}
        \includegraphics[width=\textwidth]{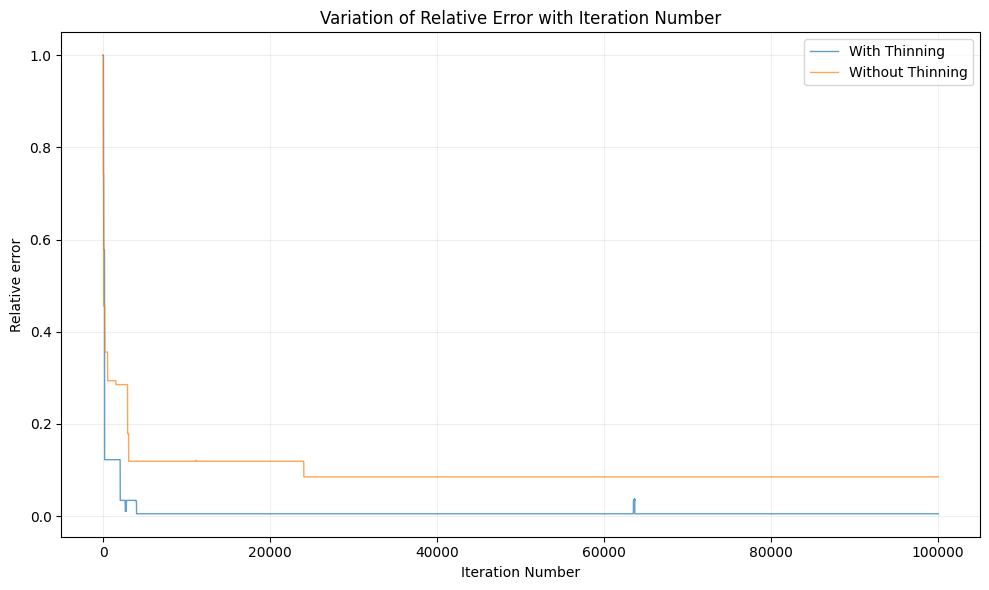}
        \caption{$N=2$}
        \label{fig6_b}
    \end{subfigure}
    \caption{The relative error trace plots.}
    \label{fig6}
\end{figure}

\begin{example}\label{exam2}
We then examine the case where data is collected  at two observation locations for fixed time point $t=1$ with  the point source intensity held constant at $1$.  
\end{example}
The reconstruction results generated by the proposed algorithm are presented in Fig.~\ref{fig4}. To examine the role of the thinning process, we conduct a comparative experiment by removing it and relying solely on the level set pCN iteration. The reconstruction results are displayed in Fig.~\ref{fig5} for $N=1$ and $N=2$. And the corresponding relative error traces are plotted in Fig.~\ref{fig6}. Obviously, when remove the thinning step in the iteration, the reconstructions are not satisfactory. This implies that the thinning process plays a important role in our algorithm. 

\begin{figure}[htbp]
    \centering
    \begin{subfigure}[b]{0.2\textwidth}
        \includegraphics[width=\textwidth]{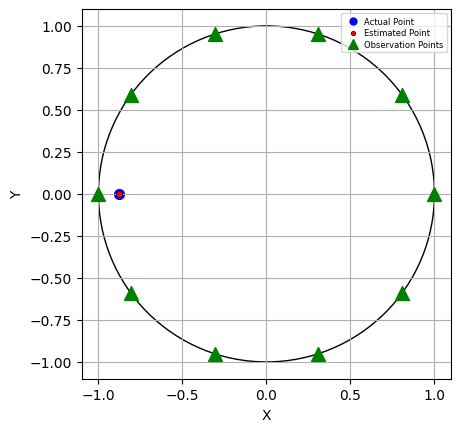}
        \caption{$N=1$}
        \label{fig2_a}
    \end{subfigure}
    \hspace{0.02\textwidth}
    \begin{subfigure}[b]{0.2\textwidth}
        \includegraphics[width=\textwidth]{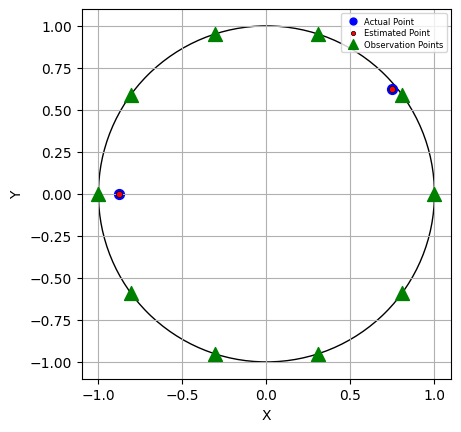}
        \caption{$N=2$}
        \label{fig2_b}
    \end{subfigure}
    \hspace{0.02\textwidth}
    \begin{subfigure}[b]{0.2\textwidth}
        \includegraphics[width=\textwidth]{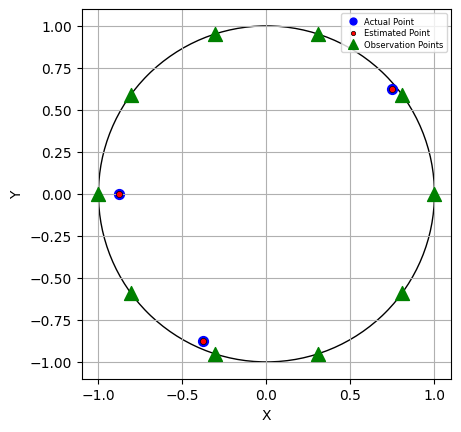}
        \caption{$N=3$}
        \label{fig2_c}
    \end{subfigure}
    \hspace{0.02\textwidth}
    \begin{subfigure}[b]{0.2\textwidth}
        \includegraphics[width=\textwidth]{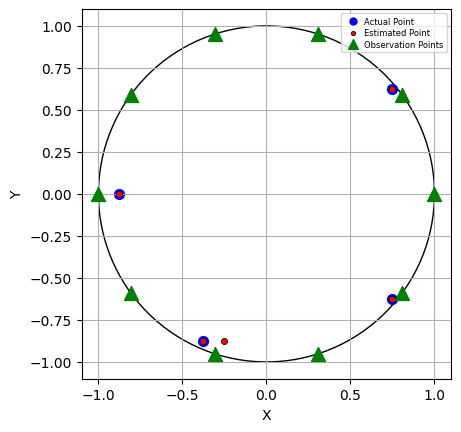}
        \caption{$N=4$}
        \label{fig2_d}
    \end{subfigure}
    \caption{The reconstruction results for point sources with different intensities.}
    \label{fig7}
\end{figure}

\begin{example}\label{exam2}
Finally, we consider the case of the point sources with different intensities using observational data at  fixed time point. 
\end{example}
Data are collected at 10 boundary observation locations (Fig.~\ref{fig7}). The reconstructed source positions for $N = 1, 2, 3, 4$ are presented in the same figure, with detailed quantitative results summarized in Table~\ref{table_a}. The results show that our method is accurate and robust. 

\begin{table}[htbp]
\centering
\captionsetup{labelfont=bf,labelsep=space,singlelinecheck=false}
\caption{Reconstruction results for point sources with varying intensities.}

\small
\resizebox{0.46\textwidth}{!}{
\begin{tabular}{c|ll|ll}
\hline
\multirow{2}{*}{$N$} & \multicolumn{2}{c|}{Exact source} & \multicolumn{2}{c}{Reconstruction} \\
\cline{2-5}
 & Position & Intensity & Position & Intensity \\
\hline
1 & $(-0.875, 0.0)$ & 0.7 & $(-0.875, 0.0)$ & 0.7024 \\
\hline
2 & $(-0.875, 0)$ & 0.7 & $(-0.875, 0.0)$ & 0.6961 \\
   & $(0.75, 0.625)$ & 0.5 & $(0.75, 0.625)$ & 0.5015 \\
\hline
3 & $(-0.875, 0.0)$ & 0.7 & $(-0.875, 0.0)$ & 0.7017 \\
   & $(0.75, 0.625)$ & 0.5 & $(0.75, 0.625)$ & 0.5003 \\
   & $(-0.375, -0.875)$ & 0.4 & $(-0.375, -0.875)$ & 0.3959 \\
\hline
4 & $(-0.875, 0)$ & 0.7 & $(-0.875, 0.0)$ & 0.6999 \\
   & $(0.75, 0.625)$ & 0.5 & $(0.75, 0.625)$ & 0.4955 \\
   & $(-0.375, -0.875)$ & 0.4 & $(-0.375, -0.875)$ & 0.2416 \\
   & $(0.75, -0.625)$ & 0.6 & $(0.75, -0.625)$ & 0.5995 \\
   & & & $(-0.25, -0.875)$ & 0.0956 \\
\hline
\end{tabular}
}
\label{table_a}
\end{table}


\section{Conclusion}
This work presents a Bayesian level set approach combined with a thinning procedure of a PPP for solving the inverse problem of reconstructing point sources in the heat equation from boundary flux measurements. The proposed method effectively integrates the level set representation for geometric flexibility and the PPP thinning process for stochastic sampling, providing a robust framework that successfully handles both the number and the intensities of the unknown sources. Numerical experiments demonstrate that the approach is both accurate and robust, yielding reliable reconstructions under various configurations, including single and multiple sources with different intensities.  Future work will extend the approach to more complex scenarios, including moving sources, time-dependent configurations, and applications in other partial differential equation models.

\bibliographystyle{cas-model2-names}

\bibliography{ref}

\begin{thebibliography}{21}
\expandafter\ifx\csname natexlab\endcsname\relax\def\natexlab#1{#1}\fi
\providecommand{\url}[1]{\texttt{#1}}
\providecommand{\href}[2]{#2}
\providecommand{\path}[1]{#1}
\providecommand{\DOIprefix}{doi:}
\providecommand{\ArXivprefix}{arXiv:}
\providecommand{\URLprefix}{URL: }
\providecommand{\Pubmedprefix}{pmid:}
\providecommand{\doi}[1]{\href{http://dx.doi.org/#1}{\path{#1}}}
\providecommand{\Pubmed}[1]{\href{pmid:#1}{\path{#1}}}
\providecommand{\bibinfo}[2]{#2}
\ifx\xfnm\relax \def\xfnm[#1]{\unskip,\space#1}\fi
\bibitem[{Badia et~al.(2005)Badia, Duong and Hamdi}]{Badia2005}
\bibinfo{author}{Badia, A.E.}, \bibinfo{author}{Duong, T.H.},
  \bibinfo{author}{Hamdi, A.}, \bibinfo{year}{2005}.
\newblock \bibinfo{title}{Identification of a point source in a linear
  advection-dispersion-reaction equation: Application to a pollution source
  problem}.
\newblock \bibinfo{journal}{Inverse Problems} \bibinfo{volume}{21},
  \bibinfo{pages}{1121--1136}.
\bibitem[{Badia and Ha-Duong(2000)}]{Badia2000}
\bibinfo{author}{Badia, A.E.}, \bibinfo{author}{Ha-Duong, T.},
  \bibinfo{year}{2000}.
\newblock \bibinfo{title}{An inverse source problem in potential analysis}.
\newblock \bibinfo{journal}{Inverse Problems} \bibinfo{volume}{16},
  \bibinfo{pages}{651--663}.
\bibitem[{Baratchart et~al.(2005)Baratchart, Abda, Hassen and
  Leblond}]{Baratchart2005}
\bibinfo{author}{Baratchart, L.}, \bibinfo{author}{Abda, A.B.},
  \bibinfo{author}{Hassen, F.B.}, \bibinfo{author}{Leblond, J.},
  \bibinfo{year}{2005}.
\newblock \bibinfo{title}{Recovery of pointwise sources or small inclusions in
  2d domains and rational approximation}.
\newblock \bibinfo{journal}{Inverse Problems} \bibinfo{volume}{21},
  \bibinfo{pages}{51--74}.
\bibitem[{Bruckner and Yamamoto(2000)}]{Bruckner2000}
\bibinfo{author}{Bruckner, G.}, \bibinfo{author}{Yamamoto, M.},
  \bibinfo{year}{2000}.
\newblock \bibinfo{title}{Determination of point wave sources by pointwise
  observations: stability and reconstruction}.
\newblock \bibinfo{journal}{Inverse Problems} \bibinfo{volume}{16},
  \bibinfo{pages}{723--748}.
\bibitem[{Cotter et~al.(2013)Cotter, Roberts, Stuart and White}]{Cotter2013}
\bibinfo{author}{Cotter, S.L.}, \bibinfo{author}{Roberts, G.O.},
  \bibinfo{author}{Stuart, A.M.}, \bibinfo{author}{White, D.},
  \bibinfo{year}{2013}.
\newblock \bibinfo{title}{Mcmc methods for functions: modifying old algorithms
  to make them faster}.
\newblock \bibinfo{journal}{Statistical Science} \bibinfo{volume}{28},
  \bibinfo{pages}{424--446}.
\bibitem[{Dashti and Stuart(2017)}]{Dashti2017}
\bibinfo{author}{Dashti, M.}, \bibinfo{author}{Stuart, A.M.},
  \bibinfo{year}{2017}.
\newblock \bibinfo{title}{The Bayesian Approach to Inverse Problems}.
\newblock 311-428, \bibinfo{publisher}{Springer International Publishing}.
\bibitem[{Egan and Mahoney(1972)}]{Egan1972}
\bibinfo{author}{Egan, B.A.}, \bibinfo{author}{Mahoney, J.R.},
  \bibinfo{year}{1972}.
\newblock \bibinfo{title}{Numerical modeling of advection and diffusion of
  urban area source pollutants}.
\newblock \bibinfo{journal}{Journal of Applied Meteorology and Climatology}
  \bibinfo{volume}{11}, \bibinfo{pages}{312--322}.
\bibitem[{de~Faria(2022)}]{Faria2022_s}
\bibinfo{author}{de~Faria, J.R.}, \bibinfo{year}{2022}.
\newblock \bibinfo{title}{A genetic algorithm for pointwise source
  reconstruction by the method of fundamental solutions}.
\newblock \bibinfo{journal}{Trends in Computational and Applied Mathematics}
  \bibinfo{volume}{23}, \bibinfo{pages}{401--412}.
\bibitem[{de~Faria et~al.(2022)de~Faria, Lesnic, Lima and
  Machado}]{Faria2022_c}
\bibinfo{author}{de~Faria, J.R.}, \bibinfo{author}{Lesnic, D.},
  \bibinfo{author}{Lima, R.}, \bibinfo{author}{Machado, T.J.},
  \bibinfo{year}{2022}.
\newblock \bibinfo{title}{The method of fundamental solutions for pointwise
  source reconstruction}.
\newblock \bibinfo{journal}{Computers and Mathematics with Applications}
  \bibinfo{volume}{114}, \bibinfo{pages}{171--179}.
\bibitem[{Gu et~al.()Gu, Zhang and Zhang}]{Gu2025}
\bibinfo{author}{Gu, Q.}, \bibinfo{author}{Zhang, W.}, \bibinfo{author}{Zhang,
  Z.}, .
\newblock \bibinfo{title}{Determine the point source of the heat equation with
  sparse boundary measurements}.
\newblock \bibinfo{journal}{https://arxiv.org/abs/2502.03018} .
\bibitem[{Huang et~al.(2021)Huang, Deng and Xu}]{Huang2021}
\bibinfo{author}{Huang, J.}, \bibinfo{author}{Deng, Z.}, \bibinfo{author}{Xu,
  L.}, \bibinfo{year}{2021}.
\newblock \bibinfo{title}{A bayesian level set method for an inverse medium
  scattering problem in acoustics}.
\newblock \bibinfo{journal}{Inverse Problems and Imaging} \bibinfo{volume}{15},
  \bibinfo{pages}{1077--1097}.
\bibitem[{Iglesias et~al.(2016)Iglesias, Lu and Stuart}]{Iglesias2016}
\bibinfo{author}{Iglesias, M.A.}, \bibinfo{author}{Lu, Y.},
  \bibinfo{author}{Stuart, A.M.}, \bibinfo{year}{2016}.
\newblock \bibinfo{title}{A bayesian level set method for geometric inverse
  problems}.
\newblock \bibinfo{journal}{Interfaces and Free Boundaries}
  \bibinfo{volume}{18}, \bibinfo{pages}{181--217}.
\bibitem[{Kovalets et~al.(2011)Kovalets, Andronopoulos, Venetsanos and
  Bartzis}]{Kovalets2011}
\bibinfo{author}{Kovalets, I.V.}, \bibinfo{author}{Andronopoulos, S.},
  \bibinfo{author}{Venetsanos, A.G.}, \bibinfo{author}{Bartzis, J.G.},
  \bibinfo{year}{2011}.
\newblock \bibinfo{title}{Identification of strength and location of stationary
  point source of atmospheric pollutant in urban conditions using computational
  fluid dynamics model}.
\newblock \bibinfo{journal}{Mathemtics and Computers in Simulation}
  \bibinfo{volume}{82}, \bibinfo{pages}{244--257}.
\bibitem[{Latz(2020)}]{Latz2020}
\bibinfo{author}{Latz, J.}, \bibinfo{year}{2020}.
\newblock \bibinfo{title}{On the well-posedness of bayesian inverse problems}.
\newblock \bibinfo{journal}{SIAM/ASA Journal on Uncertainty Quantification}
  \bibinfo{volume}{8}, \bibinfo{pages}{451--482}.
\bibitem[{Latz(2023)}]{Latz2023}
\bibinfo{author}{Latz, J.}, \bibinfo{year}{2023}.
\newblock \bibinfo{title}{Bayesian inverse problems are usually well-posed}.
\newblock \bibinfo{journal}{SIAM Review} \bibinfo{volume}{65},
  \bibinfo{pages}{831--865}.
\bibitem[{Ling and Takeuchi(2009)}]{Ling2009}
\bibinfo{author}{Ling, L.}, \bibinfo{author}{Takeuchi, T.},
  \bibinfo{year}{2009}.
\newblock \bibinfo{title}{Point sources identification problems for heat
  equations}.
\newblock \bibinfo{journal}{Communications in Computational Physics}
  \bibinfo{volume}{5}, \bibinfo{pages}{897--913}.
\bibitem[{Ling et~al.(2006)Ling, Yamamoto, Hon and Takeuchi}]{Ling2006}
\bibinfo{author}{Ling, L.}, \bibinfo{author}{Yamamoto, M.},
  \bibinfo{author}{Hon, Y.C.}, \bibinfo{author}{Takeuchi, T.},
  \bibinfo{year}{2006}.
\newblock \bibinfo{title}{Identification of source locations in two-dimensional
  heat equations}.
\newblock \bibinfo{journal}{Inverse Problems} \bibinfo{volume}{22},
  \bibinfo{pages}{1289--1305}.
\bibitem[{Reich and Weissmann(2021)}]{Reich2021}
\bibinfo{author}{Reich, S.}, \bibinfo{author}{Weissmann, S.},
  \bibinfo{year}{2021}.
\newblock \bibinfo{title}{Fokker-planck particle systems for bayesian
  inference: computational approches}.
\newblock \bibinfo{journal}{SIAM/ASA Journal on Uncertainty Quantification}
  \bibinfo{volume}{9}, \bibinfo{pages}{446--482}.
\bibitem[{Ren and Zhong(2019)}]{Ren2019}
\bibinfo{author}{Ren, K.}, \bibinfo{author}{Zhong, Y.}, \bibinfo{year}{2019}.
\newblock \bibinfo{title}{Imaging point sources in heterogeneous enviroments}.
\newblock \bibinfo{journal}{arXiv:1901.07189v2}
  \bibinfo{volume}{https://arxiv.org/pdf/1901.07189}.
\bibitem[{Stuart(2010)}]{Stuart2010}
\bibinfo{author}{Stuart, A.M.}, \bibinfo{year}{2010}.
\newblock \bibinfo{title}{Inverse problems: A bayesian perspective}.
\newblock \bibinfo{journal}{Acta Numerica} \bibinfo{volume}{19},
  \bibinfo{pages}{451--559}.
\bibitem[{Zhang et~al.(2024)Zhang, Chang and Guo}]{Zhang2024}
\bibinfo{author}{Zhang, D.}, \bibinfo{author}{Chang, Y.}, \bibinfo{author}{Guo,
  Y.}, \bibinfo{year}{2024}.
\newblock \bibinfo{title}{Jointly determining the point sources and obstacle
  from cauchy data}.
\newblock \bibinfo{journal}{Inverse Problems} \bibinfo{volume}{015014},
  \bibinfo{pages}{25pp}.

\end{thebibliography}

%
%
%

\end{document}